\documentclass[final,5p,times,twocolumn]{elsarticle}

\usepackage{amssymb}
\usepackage{amsmath}

\usepackage{gensymb}
\usepackage{float}
\usepackage{placeins} 
\newcommand{\figref}[1]{Fig.~\ref{#1}}

\begin{document}

\begin{frontmatter}


\title{Wrinkling in Selected Polymer Thin Films Induced by Combined Ion Beam and Humidity Exposure}

\author[label_a]{Alessia Danagoulian\fnref{label1}} 
\fntext[label1]{These authors contributed equally to this work as co-first authors.}
\affiliation[label_a]{organization={Department of Physics, Boston University},
            city={Boston},
            postcode={02215}, 
            state={Massachusetts},
            country={USA}}

\author[label_b]{Benli Jiang\corref{cor1}\fnref{label1}} 
\ead{jbl12@bu.edu}
\cortext[cor1]{Corresponding authors:}
\affiliation[label_b]{organization={Division of Materials Science and Engineering, Boston University},
            city={Boston},
            postcode={02215}, 
            state={Massachusetts},
            country={USA}}

\author[label_a]{Nicholas Russo\corref{cor1}} 
\ead{nzr@bu.edu}

\author[label_a]{Colette Abadie} 

\author[label_c]{Jalal Karimzadeh Khoei} 
\affiliation[label_c]{organization={Faculty of Engineering and Natural Sciences,Sabancı University},
            city={Istanbul},
            postcode={34956}, 
            country={Türkiye}}

\author[label_a,label_d]{Grace Pettis} 
\affiliation[label_d]{organization={School of Chemical, Biological and Environmental Engineering, Oregon State University},
            city={Corvallis},
            postcode={97331}, 
            state={Oregon},
            country={USA}}

\author[label_a,label_e]{Jocelyn Zhang} 
\affiliation[label_e]{organization={Department of Electrical Engineering and Computer Science, Massachusetts Institute of Technology},
            city={Cambridge},
            postcode={02139}, 
            state={Massachusetts},
            country={USA}}

\author[label_a,label_f]{Neil Baker} 
\affiliation[label_f]{organization={School of Physics, Georgia Institute of Technology},
            city={Atlanta},
            postcode={30332}, 
            state={Georgia},
            country={USA}}

\author[label_c]{Eda Güney} 

\author[label_c,label_g]{Omid Moradi} 
\affiliation[label_g]{organization={Sabancı University Nanotechnology Research and Application Center},
            city={Istanbul},
            postcode={34956}, 
            country={Türkiye}}

\author[label_b]{Wei-Jing Chen} 

\author[label_b]{Jiaqi Tang} 

\author[label_a,label_h]{Robert Sims, Jr.} 
\affiliation[label_e]{organization={School of Engineering, Wentworth Institute of Technology},
            city={Boston},
            postcode={02215}, 
            state={Massachusetts},
            country={USA}}

\author[label_a,label_b,label_i]{Kevin E. Smith} 
\affiliation[label_i]{organization={Department of Chemistry, Boston University},
            city={Boston},
            postcode={02215}, 
            state={Massachusetts},
            country={USA}}

\author[label_c,label_g]{Gozde Ozaydin Ince\corref{cor1}} 
\ead{gozdeince@sabanciuniv.edu}

\author[label_a,label_b]{Karl F. Ludwig, Jr.} 

\begin{abstract}
This study investigates ion beam sputtering (IBS)-induced surface wrinkling phenomena in three polymers with varying hydrophilicity: poly-hydroxy-ethyl-methacrylate (pHEMA), poly-4-vinyl pyridine (p4VP), and poly-2,4,6,8-tetramethyl-2,4,6,8-tetravinylcyclotetrasiloxane (pV4D4). 
It is observed that pHEMA and p4VP films wrinkle only when exposed to ion bombardment and subsequent water vapor exposure. No wrinkling is observed in pV4D4 under these same conditions.
X-ray photoelectron spectroscopy (XPS) and Fourier transform infrared spectroscopy (FTIR) are performed before IBS, after IBS, and after exposure to humidity. XPS shows that IBS drives chemical changes within the surface layer, creating a graphitized film at the surface.
For the polymer films that exhibit wrinkling (pHEMA and p4VP), XPS and FTIR indicate water absorption in both the surface and the bulk of the films, resulting in swelling. 
We conjecture that the formation of wrinkles arises from this swelling being mechanically constrained by the rigid underlying silicon substrate and the stiff graphitized surface layer. 
In contrast, the absence of wrinkle formation in pV4D4 under the same experimental conditions can be attributed to its comparatively low water absorption and the correspondingly limited swelling response.
\end{abstract}


\end{frontmatter}


\section{Introduction}
\label{sec:intro}

Polymer films are ubiquitous in scientific and engineering applications, and the optimization of polymer film morphology has gained increasing attention, particularly with respect to the formation and manipulation of wrinkling phenomena. Wrinkle formation has multiple uses, including increasing surface area for catalysis or adsorptive storage, and for micro-structural engineering in optical devices and photonic media, such as polarizers, filters, and lenses \cite{liuLSmartWrinkledInterfaces2024, enrightAFMSelfWrinklingVaporDepositedPolymer2022, chenPIWrinklingInstabilitiesPolymer2012}.
Whether wrinkling is desirable in a particular situation depends on the target application; therefore, it is an important factor to consider when choosing or designing polymers and polymer-based heterostructures \cite{chenPIWrinklingInstabilitiesPolymer2012,hendricksNLWrinkleFreeNanomechanicalFilm2007}. 
Wrinkling is a mechanically governed phenomenon that depends on the elastic properties and microstructural characteristics of the film, with wrinkle formation arising from the relaxation of externally induced stresses \cite{Cerda2003}.
These stresses, however, may be driven by non-mechanical factors, such as temperature gradients, changes in material chemistry, or solvent interactions \cite{chenSMWrinklingInhomogeneouslyStrained2013,sharpPREMechanicallyDrivenWrinkling2007}.
The control of wrinkling via adjustable experimental parameters, such as film depth, materials chemistry, and the manner in which stress is applied, is of interest. 

Wrinkle structures have been experimentally observed on polymer films using various approaches, including post-deposition annealing \cite{Bowden1998,Das2017,okayasuAFMSpontaneousFormationOrdered2004}, ion beam sputtering (IBS) \cite{moonPNASUWrinkledHardSkins2007,moonSMControlledFormationNanoscale2007,moonSaCTSculptingPolymersUsing2008,Jeong2015,Jeong2015Srep,jeongLTailoringOrientationPeriodicity2016,Arias2021}, UV-ozone treatment \cite{Breid2011}, and plasma treatment \cite{Chua2000,ahmadAFMSurfaceWrinklingPlasmaExposed2025}. Each of these approaches requires the introduction of a surface layer whose mechanical properties differ from those of the underlying material. In IBS, UV-ozone, and plasma treatment approaches, this surface layer is generated through the chemical modification of the outermost region of materials that is directly exposed to the ion beam, UV-ozone, or plasma. 

In the case of IBS of poly-dimethylsiloxane (PDMS), it has been proposed that wrinkling is caused by the creation of a stiff surface layer and the swelling of the underlying bulk polymer, both induced by the IBS. Specifically, collisions between the ions and polymer molecules beneath the surface induce thermal expansion that is constrained by the stiff surface layer. This causes the surface to buckle \cite{moonPNASUWrinkledHardSkins2007,Jeong2015,Jeong2015Srep,jeongLTailoringOrientationPeriodicity2016}. Recently, Ahmad \textit{et al}. \cite{ahmadAFMSurfaceWrinklingPlasmaExposed2025} experimentally observed that PDMS requires both plasma and water vapor exposure to form wrinkles, suggesting that swelling due to water vapor absorption is responsible for the wrinkling of PDMS after plasma exposure. Our study focuses on the IBS and subsequent humidity exposure of three polymer films with different chemical structure and hydrophilicity: poly-hydroxy-ethyl-methacrylate (pHEMA), poly-4-vinyl pyridine (p4VP), and poly-2,4,6,8-tetramethyl-2,4,6,8-tetravinylcyclotetrasiloxane (pV4D4). Similarly, we identify the presence of water vapor after IBS as a necessary condition for wrinkle formation in pHEMA and p4VP. Atomic force microscopy (AFM) was used to monitor the morphology, and \textit{ex situ} and \textit{in situ} X-ray photoelectron spectroscopy (XPS) were used to obtain surface and near surface chemistry—with Fourier transform infrared spectroscopy (FTIR) providing bulk chemical information. 

We demonstrate that IBS creates an effective graphitic carbon-polymer heterostructure, which is consistent with previously reported results \cite{calcagnoNIaMiPRSBBIwMaAStructuralModificationPolymer1992,goodwinCSoftIonSputtering2020,nesovPSSElectronicStructureNitrogencontaining2017}. In addition, we find that wrinkling arises in pHEMA and p4VP when swelling induced by water absorption after IBS is constrained by the stiff graphitic surface layer and the hard substrate of the polymer film. Moreover, wrinkling is strongly suppressed in pV4D4, a polymer with low water absorption and thus minimal swelling. 
These results add to the growing body of literature regarding the formation of wrinkling in polymer thin films induced by ion bombardment \cite{moonPNASUWrinkledHardSkins2007,moonSMControlledFormationNanoscale2007,moonSaCTSculptingPolymersUsing2008,Jeong2015,Jeong2015Srep,jeongLTailoringOrientationPeriodicity2016,Arias2021}.

\section{Experimental Methods}
\label{sec:expt}
\subsection{Materials and Film Synthesis}
\label{subsec:synthesis}

Monomers 2-Hydroxyethyl methacrylate (HEMA), 4-Vinyl Pyridine (4VP), 2,4,6,8-tetramethyl-2,4,6,8-tetravinylcyclotetrasiloxane (V4D4), and the initiator tert-butyl peroxide (TBPO) were purchased from Sigma and used without further purification.
\begin{figure}[ht!]
    \centering
    \includegraphics[width=\linewidth]{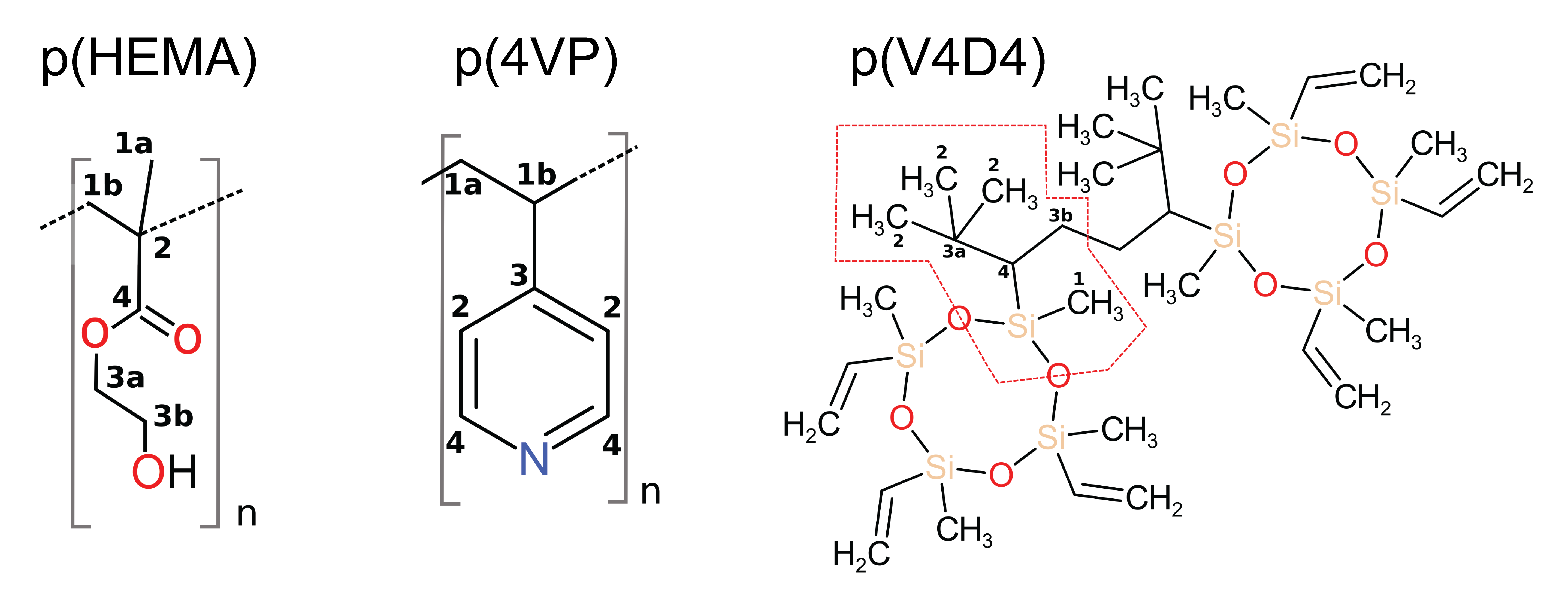}
    \caption{Polymeric units of the films used in this study.
    The peak assignments in the XPS spectra shown in \figref{fig:Total_C1sXPS} correspond to the numbering for each unit, where atoms whose peaks overlap are labeled `1a, 1b...'. Atoms that are symmetry equivalent have the same label. A selection of possible contributions to the C 1s XPS from polymerized V4D4 is circled in red. }
    \label{fig:Molecules}
\end{figure}

Polymer thin films were deposited on silicon wafers using initiated chemical vapor deposition (iCVD) in a vacuum reactor.
The flow rates of monomers were regulated with metering needle valves (H1300 series, Ham-Let Group, USA). TBPO was delivered through a mass-flow controller (GFC17, Aalborg Instruments \& Controls Inc., USA) at room temperature. 
The stage temperature of the reactor was controlled using a recirculating chiller (SM3, Labo Inc., Turkey). 
For p4VP deposition, 4-vinylpyridine (4VP) vapor was introduced at 1 sccm from a monomer jar maintained at 60°C, while TBPO was introduced at 1 sccm (room temperature).
The substrate temperature was fixed at 12°C, and the filament temperature was set to 250°C. For pHEMA deposition, hydroxyethyl methacrylate (HEMA) vapor was introduced at 1.5 sccm from a monomer jar maintained at 80°C, while TBPO was introduced at 1 sccm (room temperature). 
The substrate temperature was maintained at 20°C, and the filament temperature was set to 250°C. For pV4D4 film deposition, the V4D4 vapor was introduced at 2 sccm from a monomer jar maintained at 85°C, while TBPO was introduced at 1 sccm (room temperature). 
The substrate temperature was maintained at 30°C, and the filament temperature was set to 250°C. The thickness of all polymer films on Si was in the range of 250-300 nm, measured by ellipsometry.

\subsection{Ion Beam Sputtering and Post-Sputtering Sample Handling}
\label{subsec:IBS}
\begin{figure*}[h]
    \centering
    \includegraphics[width=0.8\linewidth]{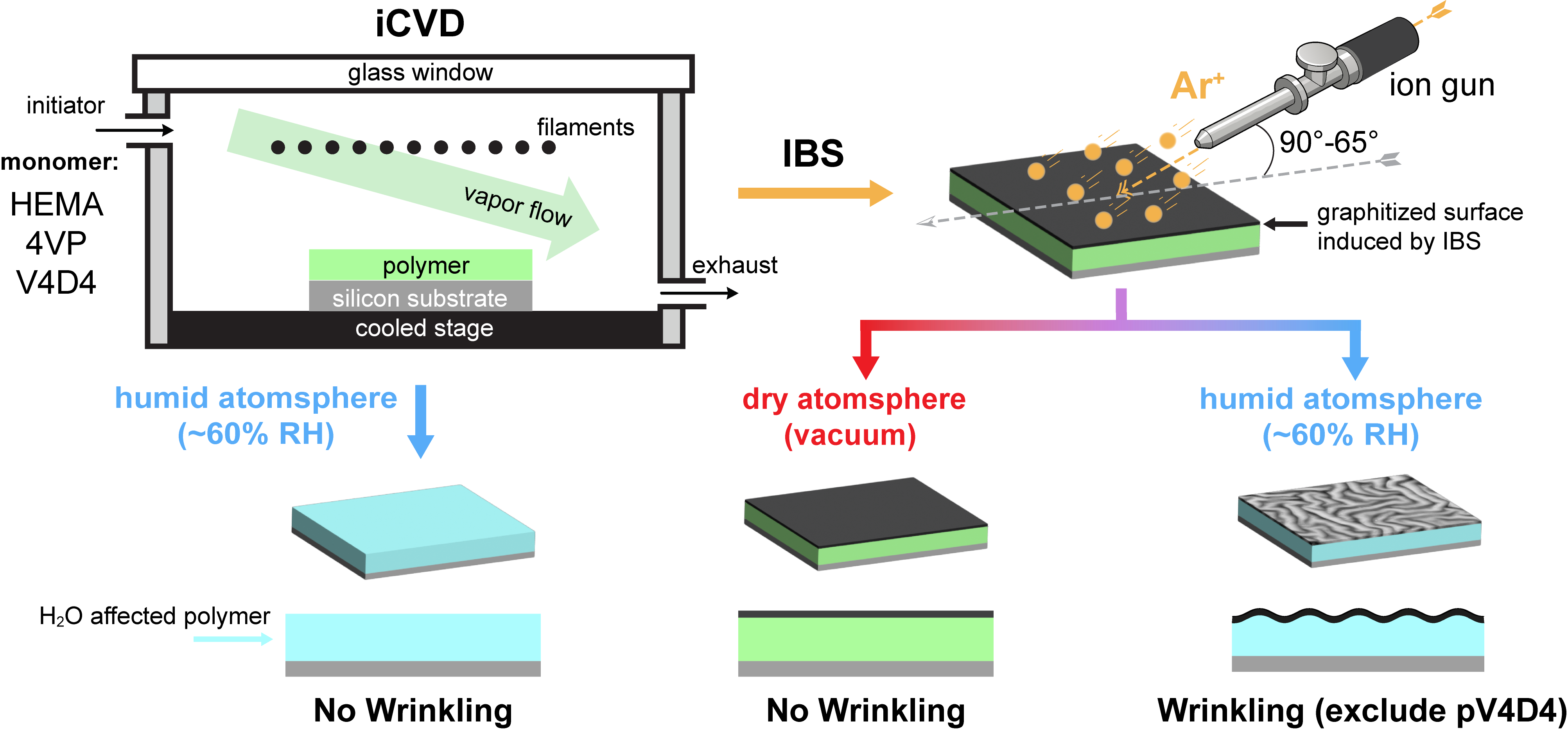}
    \caption{Schematic of the wrinkling in the pHEMA and p4VP films in the humid atomsphere after IBS. In contrast, in the dry atmosphere after IBS or in the humid atmosphere without IBS, all three polymers studied (pHEMA, p4VP and pV4D4) remain flat with no wrinkles observed.}
    \label{fig:schematics}
\end{figure*}

IBS of polymer samples was conducted in a custom ultra-high vacuum chamber with a base pressure of $1 \times 10^{-7} \text{ Torr}$. Polymer thin film samples were bombarded at a polar angle of 65° with monoatomic $\mathrm{Ar}^+$ ions at room temperature, as shown in \figref{fig:schematics}. 
To prevent charge accumulation and adequately control the sample temperature during IBS, vacuum compatible thermal contact silver paste (MeiVac TP-832) was applied to the entire back of the samples, and they were mounted flat on the grounded stainless steel sample platen. To exclude potential influence from IBS-induced heating, sacrificial samples were prepared using the same procedures and IBS conditions described in this section, with an additional non-reversible thermometer sticker affixed to the back of the samples, where it is shielded from direct ion irradiation. As measured by the thermometer sticker, the sample temperature did not exceed 50°C during IBS, indicating that thermal effects were minimized.
The ion beam was generated by a Nonsequitur Technologies (NTI) 1402 ion gun; during the IBS treatment, the ion gun was backfilled with UHP Argon, and the chamber was kept at an operating pressure of $4 \times 10^{-6} \text{ Torr}$ due to the differential pumping design of the NTI 1402 ion gun. 
The $\text{Ar}^+$ ion beam energy was 2000 eV. The ion beam raster area was 1$\times$1 cm$^2$, and the average ion beam flux over the raster area was $2.0\times 10^{13}$ ions $\text{cm}^{-2}$ $\text{s}^{-1}$. The final ion fluence was $1.2\times 10^{16}$ ions $\text{cm}^{-2}$ for 10 minutes of IBS. In addition, some samples were bombarded \textit{in situ} in the XPS system, as described in Section \ref{subsec:XPS-method}.

In order to investigate the effect of humidity, post sputtering samples were held in either a humid or a dry environment. 
The controlled humid environment was created in an acrylic container with distilled water at the bottom. A hot plate beneath the container was kept at 60°C to maintain the temperature of the distilled water and produce water vapor to interact with the sample. 
The enclosure was partially sealed to provide sufficient ventilation and avoid saturation. The samples were placed on an elevated rack inside the container, separate from the water at the bottom.
A DHT11 sensor was placed in the container to record the temperature and relative humidity near the samples at 1-hour intervals. 
The temperature near the samples for the humid environment study was maintained at 23°C $\pm$ 2°C, with a humidity of 60\% RH $\pm$ 10\%. 
For comparison, a dry environment was created in a sealed stainless steel vacuum chamber at room temperature; the chamber was evacuated and backfilled with UHP Argon to a pressure of approximately 10 Torr. The pressure was kept below ambient pressure for achieving the sealing with the O-rings in the vacuum chamber, which prevents moisture permeation from the ambient environment.
During sample transfer and AFM imaging, room temperature and humidity were measured at 25°C $\pm$ 2°C and 20\% RH $\pm$ 5\%.

\subsection{Surface Characterization}

\subsubsection{Surface Morphology}

The surface morphology of each sample was investigated using a Bruker MultiMode V AFM with a PPP-NCHR probe from Nanosensors. The surface root mean square (RMS) roughness and 2-dimensional (2D) power spectral density (PSD) analysis were acquired using Python, following the analysis principles of Gwyddion software \cite{necasOPGwyddionOpensourceSoftware2012}.
Since the wrinkling patterns are isotropic, the wavelength of the wrinkles was determined through the peak position in the radial PSD analysis, which is derived from the 2D PSD analysis. The details of the wavelength determination are presented in \ref{subsec:AFM-analysis}.

\subsubsection{Infrared Absorption Spectroscopy}
 
Fourier Transform Infrared (FTIR) absorption spectroscopy measurements were taken in attenuated total reflectance (ATR) mode with a ZnSe crystal using a Nicolet Nexus 670 spectrometer.
The samples were scanned from 1000 to 4000 cm$^{-1}$ at a resolution of 2 cm$^{-1}$ with 32 scans.
The probe depth of ATR-FTIR is estimated to be up to several microns, much greater than the film thickness \cite{kaneJBMRBABATRFTIRThicknessMeasurement2009, finaASRDepthProfilingPolymer1994, kaurMToSFundamentalsATRFTIRSpectroscopy2021, gotzSAPAMaBSApparentPenetrationDepth2020, larocheASSFTIRATRSpectroscopyThin2013}.

\subsubsection{X-ray Photoelectron Spectroscopy (XPS)}\label{subsec:XPS-method}

XPS measurements were conducted using a PHI Genesis XPS system with monochromated Al K$\alpha$ radiation. 
A set of \textit{ex situ} XPS measurements was obtained before ion bombardment, immediately after bombardment, and after both bombardment and exposure to the aforementioned humid environment, in order to analyze changes in the surface composition of samples.
The X-ray beam spot size was 100 microns, with a beam power of 25 W @ 15 kV as set by the manufacturer.
No changes in the spectra due to radiation damage by the X-ray beam on the polymers were observed.

For each polymer, the C 1s, O 1s, and survey spectra were taken. In addition, N 1s and Si 2p spectra were collected for p4VP and pV4D4, respectively. The survey spectra used a pass energy of 114 eV, while the higher resolution scans around individual peaks used a pass energy of 27 eV.

A set of \textit{in situ} XPS measurements using the Argon ion gun (ion beam spot size 0.5 mm in diameter, raster area 3$\times$3 mm$^2$) installed on the XPS instrument were also taken, using an $\text{Ar}^+$ ion energy of 2 keV. The ion beam current was 347 nA, as specified by the manufacturer, corresponding to an average ion flux over the raster area of \(2.4\times 10^{13}\) ions cm\(^{-2}\) s\(^{-1}\). This flux is comparable to that provided by the NTI ion gun for all \textit{ex situ} experiments discussed in Section \ref{subsec:IBS}.  
No Zalar rotation was used during bombardment, and for these measurements, a pass energy of 55 eV was used for the XPS elemental spectra.
A series of XPS elemental spectra were collected after each ion beam sputtering step. The duration of each sputtering step was 6 s. The \textit{in situ} XPS measurements thus tracked the evolution of the surface chemical structure over 16 sputtering steps, corresponding to a total ion sputtering time of 96 s.
The angle between the ion beam and the sample at the X-ray focus spot was 55°.
Quantitative analysis was performed using the corrected empirical relative sensitivity factors (ceRSFs) taken from the MultiPAK software on the instrument, and the equivalent homogeneous fraction was calculated. 
For all cases, the U 3 Tougaard background was used to analyze the C 1s spectra \cite{tougaardJoVS&TAPracticalGuideUse2020}. 
The peak models used in the analysis, specific to each polymer film under study, are provided in Table \ref{tab:XPS_Lineshapes} in \ref{subsec:XPS-analysis}.

\section{Results}

\begin{figure*}[h!]
    \centering
    \includegraphics[width=0.8\linewidth]{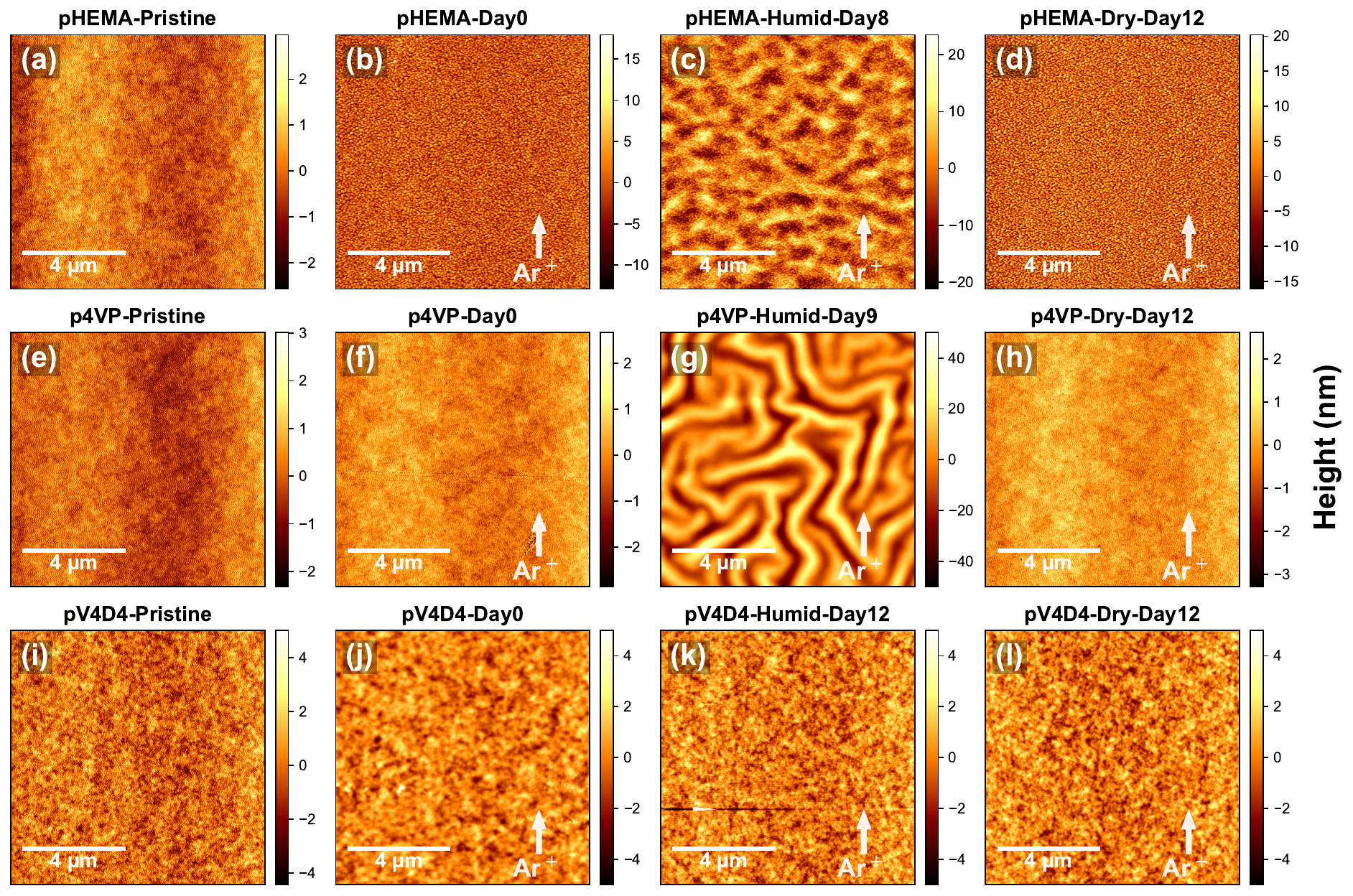}
    \caption{AFM images of polymer surfaces before ion bombardment (a,e,i); immediately after ion bombardment (b,f,j); after exposure to a humid atmosphere post-bombardment (c,g,k); and after exposure to a dry atmosphere post-bombardment (d,h,l). Row 1 (a-d) corresponds to a pHEMA film right after bombardment (b), after 8 days in the humid atmosphere (c), and 12 days in the dry atmosphere (d). Row 2 (e-h) corresponds to a p4VP film right after bombardment (f), after 9 days in the humid atmosphere (g) and 12 days in the dry atmosphere (h). Row 3 (i-l) corresponds to a pV4D4 film right after bombardment (j), after 12 days in the humid atmosphere (k), and 12 days in the dry atmosphere (l). The white arrow indicates the direction of ion beam incidence.}
    \label{fig:AFM_Matrix}
\end{figure*}

\subsection{Changes in Film Morphology}

\begin{figure*}[h]
    \centering
    \includegraphics[width=0.85\linewidth]{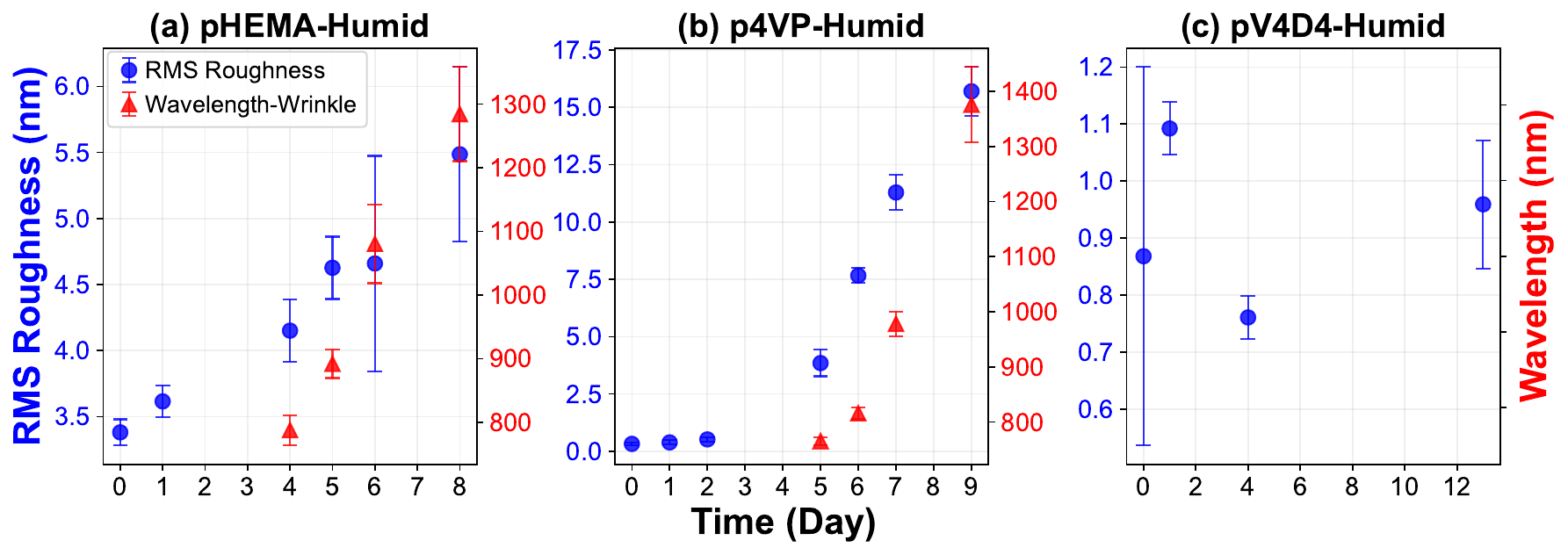}
    \caption{AFM analysis of surface roughness and wrinkle wavelength as a function of time exposed to a humid atmosphere after IBS of pHEMA (a), p4VP (b) and pV4D4 (c). For pV4D4, no wrinkles were observed through the 12 days post IBS.}
    \label{fig:AFM_Analysis}
\end{figure*}

The AFM images of the deposited films after the different treatment steps are shown in Fig. \ref{fig:AFM_Matrix}. 
Immediately after the initial IBS treatment, none of the films exhibit wrinkling, as shown in panels (b), (f), and (j). There is also minimal change in the roughness of the surface after multiple days in the dry environment, as shown in panels (d), (h), and (l). Unbombarded films show no change in surface morphology if they are exposed to the humid environment for 12 days (not shown).

For ion bombarded pHEMA and p4VP, however, a significant wrinkling pattern forms after exposure to water vapor in the humid environment, as seen in panels (c) and (g) of Fig. \ref{fig:AFM_Matrix}, and also in Fig.~\ref{fig:AFM_HEMA} and~\ref{fig:AFM_4VP} in \ref{subsec:AFM-analysis}. The insets in Fig.~\ref{fig:AFM_HEMA} and~\ref{fig:AFM_4VP} demonstrate a pronounced circular feature at the center of the 2D PSD, indicating that the wrinkle pattern is isotropic. Of the polymers studied, pHEMA and p4VP developed tall, well-defined wrinkles, with amplitudes of nearly 25 nm (pHEMA) and 40 nm (p4VP) within 8 to 9 days inside the humid environment. 
In contrast, pV4D4 remained flat, with no noticeable wrinkles in the humid environment after 12 days post-IBS. The changes in root mean squared (RMS) surface roughness and the wrinkling wavelength for each polymer are plotted over time in \figref{fig:AFM_Analysis}. 
The RMS surface roughness is assumed to be directly proportional to the mean wrinkle amplitude.
Both pHEMA and p4VP exhibit clear wrinkle formation, with each of their characteristic wavelengths and RMS surface roughness increasing with water vapor exposure time.

In comparison to the continuous formation and coarsening process observed in pHEMA, there is a delayed onset of the formation of wrinkles in p4VP. p4VP also demonstrates visibly greater pattern coherence than pHEMA. The p4VP sample (Fig. \ref{fig:AFM_Matrix}(g)) exhibits smoothly curving ridges with a quasi-sinusoidal height profile (Fig. \ref{fig:AFM_4VP}), as well as a larger amplitude and correlation length. Similarly, pHEMA (Fig. \ref{fig:AFM_Matrix}(c)) displays an isotropic pattern and maintains a characteristic wavelength but contrastingly demonstrates shorter, disordered ridges with a noisy height profile.  

Notably, as seen in Fig. \ref{fig:AFM_Matrix}(k), pV4D4 does not experience any wrinkling formation even over a longer 12 day time period of water vapor exposure.

These results suggest that the observed wrinkling phenomenon depends on the intrinsic chemical properties of the polymer, IBS, and the subsequent humidification process. To further study changes in the polymer film chemistry, \textit{ex situ} XPS was employed.

\subsection{Changes in Film Surface Chemistry}

\begin{figure*}[th!]
    \centering
    \includegraphics[width=.95\linewidth]{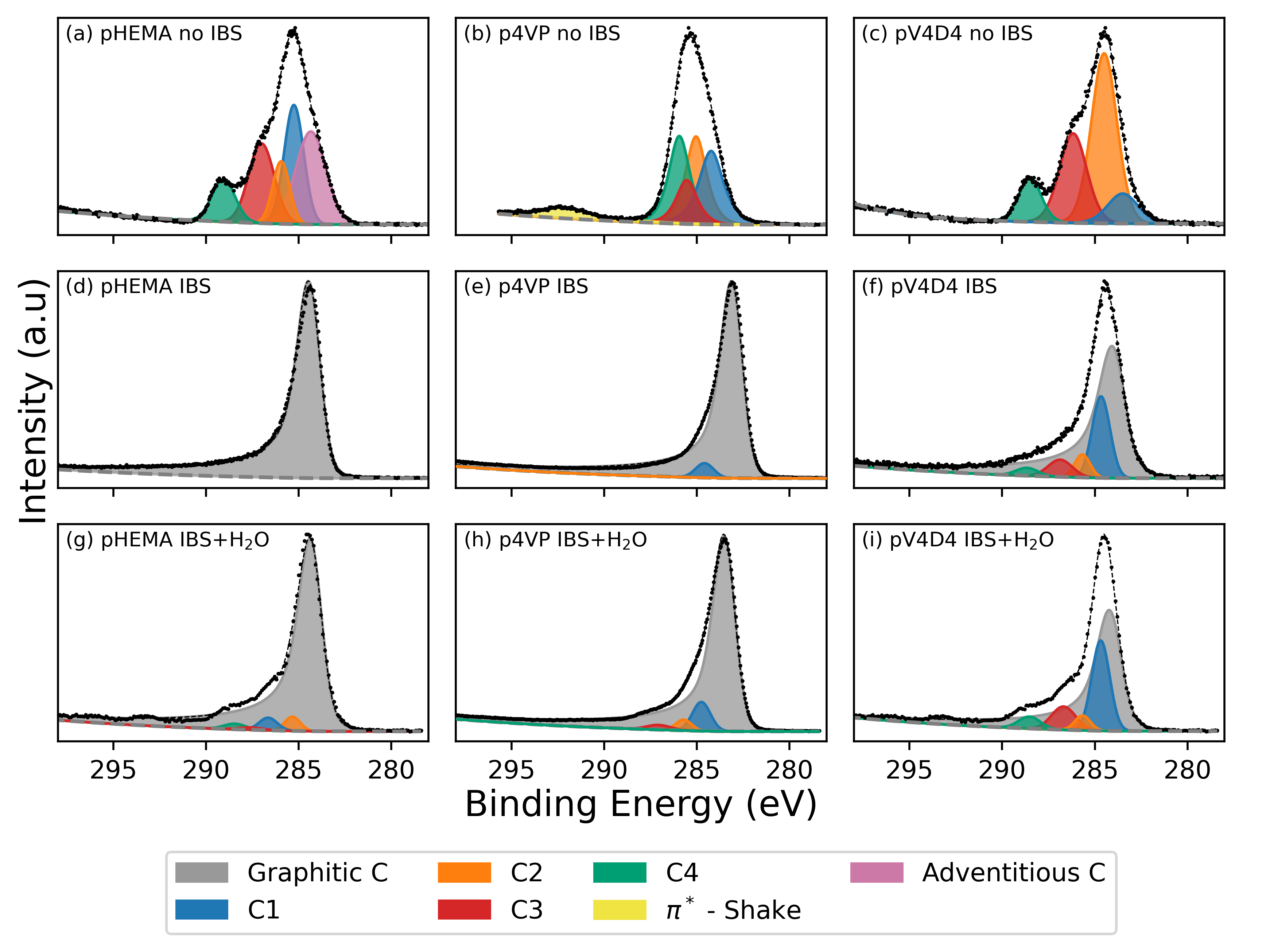}
    \caption{ C 1s XPS of pHEMA (a,d,g), p4VP (b,e,h), and pV4D4 (c,f,i) polymer samples before sputtering, after sputtering, and after water exposure. Row 1 shows the material pre-IBS bombardment. Row 2 shows the material post IBS but before exposure to the humid environment. Row 3 shows the polymer species after IBS and exposure to the humid environment. C1 component likely corresponds to C-H `aliphatic carbon', C2 to C-O or C-OH, C3 to C=O, and C4 to O-C=O.}
    \label{fig:Total_C1sXPS}
\end{figure*}

The C 1s XPS spectra before IBS, after IBS, and after exposure to high humidity are shown in \figref{fig:Total_C1sXPS}. 

The XPS spectra for the pHEMA sample are shown in \figref{fig:Total_C1sXPS}(a,d,g).
The initial film has strong agreement with previous references of monomer HEMA XPS when the adventitious layer component, labeled `AdC', is extracted \cite{castnerSSSCharacterizationPoly2HydroxyethylMethacrylate1996}.
After IBS, the XPS data suggest that, up to the C 1s probe information depth in this material ($\lambda_{C1s}< 10$ nm), the uppermost layer is composed entirely of graphitic or graphitized carbon. This is consistent with previous work modeling the graphitic peak as a single finite asymmetric Lorentzian function \cite{estrade-szwarckopfCXPSPhotoemissionCarbonaceous2004, blumeCCharacterizingGraphiticCarbon2015,al-gaashaniCIXPSStructuralStudies2019b}.
After exposure to water vapor, the graphitized film acquires additional hydro-carbon and oxidized carbon features; these features are at varied binding energies of XPS and depend strongly on the chemical environment to which the sample is exposed. 
These features are difficult to assign definitively, but they are consistent with the findings of Chen \textit{et al}. \cite{afanasevBComparativeInvestigationXPS2023, chenFNaCNReviewC1sXPSspectra2020}, who suggest that in graphitic carbon systems which have some partial oxidation, the C-H has a binding energy near 285 eV, while C-O, C=O, and O-C=O features may be present in wide ranges from 286 to 289 eV. The peak assignment in our study follows this framework that in \figref{fig:Total_C1sXPS}(g) the C1 component likely corresponds to C-H `aliphatic carbon', C2 to C-O or C-OH, C3 to C=O, and C4 to O-C=O. This suggests water absorption and oxidation of the film surface, resulting in the formation of oxidized surface graphitic groups.

In \figref{fig:Total_C1sXPS}(a), the relative peak areas do not satisfy the expected peak area ratio for pHEMA. It is clear that the C 1s in (g) cannot be represented as a sum of the graphitic phase and pHEMA.
This suggests that these peaks correspond to the functionalization of the surface after the interaction with water \cite{tougaardJoVS&TAPracticalGuideUse2020, baerJoVS&TAWhatMoreCan2025}.

\begin{figure}
    \centering
    \includegraphics[width=\linewidth]{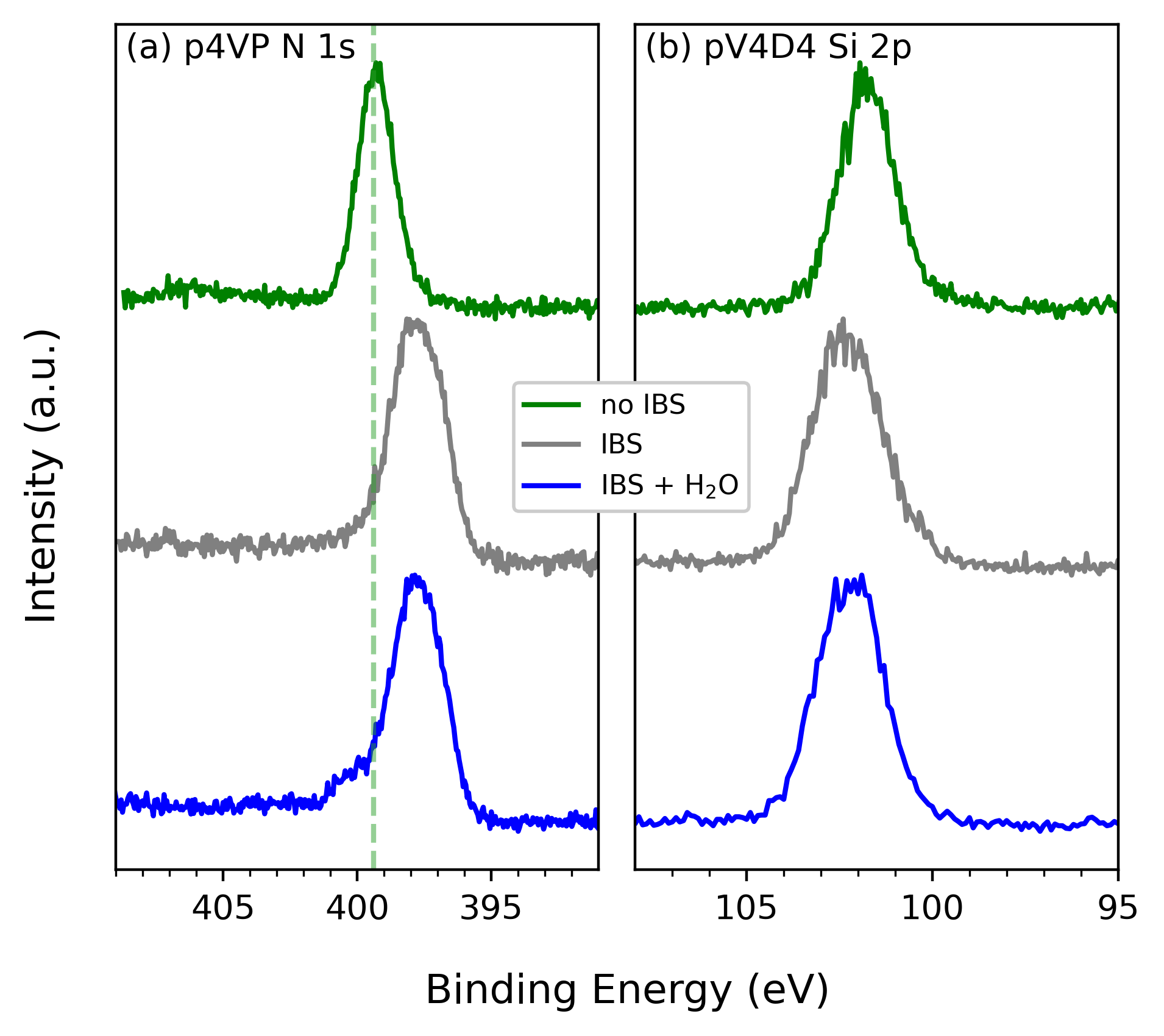}
    \caption{(a) N 1s XPS of p4VP polymer sample before sputtering, after sputtering, and after water vapor exposure. (b)  Si 2p XPS of pV4D4 polymer sample before sputtering, after sputtering, and after water vapor exposure.}
    \label{fig:ExtraXPS_SiN}
\end{figure}

Presented in \figref{fig:Total_C1sXPS}(b,e,h) are the XPS results for p4VP. 
There is a complete disappearance of the peak components assigned to C-N species and carbon atoms adjacent to the pyridine nitrogen in the 286 eV region, together with the pyridinic nitrogen feature and the associated aromatic $\pi$-$\pi^*$ shake-up satellite near 295 eV. This suggests substantial degradation or disruption of the pyridine ring structure in p4VP \cite{maazJMSPoly4vinylpyridinemodifiedSilicaEfficient2019,bagusJCPXPSPyridineCombined2025}.
While pHEMA is composed of C, O, and H, p4VP contains C, N, and H, with trace amounts of O either physisorbed to the surface or chemisorbed in the form of a pyridine-N-oxide group.
This results in the presence of nitrogen in the graphitized carbon phase after IBS, as shown in \figref{fig:ExtraXPS_SiN}(a).
Nitrogen is well known to be present in a variety of carbon allotropes, including graphite \cite{kuntumallaASSNitrogenBondingWork2020,kusunokiSSXPSStudyNitridation2001, titantahDaRMCarbonNitrogen1s2007, kondoPRBAtomicscaleCharacterizationNitrogendoped2012, kiuchiPCCPCharacterizationNitrogenSpecies2016,hellgrenInterpretationXrayPhotoelectron2016}.

Following \figref{fig:ExtraXPS_SiN}(a), it can be observed that there is a significant change in the N 1s spectra at each step of the treatment.
In the initial material, a single peak corresponding to the pyridinic N peak can be readily observed; after the IBS treatment, this peak and the $\pi^*$ feature disappear and are replaced by a broad peak, consistent with the presence of multiple graphene-nitrogen moieties, such as cyanidic or pyrrolic N, as discussed by Kiuchi \textit{et al}. \cite{kuntumallaASSNitrogenBondingWork2020,kusunokiSSXPSStudyNitridation2001,kiuchiPCCPCharacterizationNitrogenSpecies2016,kiuchiNRLLewisBasicityNitrogenDoped2016}.
After exposure to water, in addition to broadening of the main feature and a significant spectral shift to about 0.5 eV higher binding energy, a shoulder/peak also appears at 403 eV, which is consistent with Nitrous oxide (NO).

These findings are consistent with previous literature on ion beam damage induced by IBS treatment of polymer films, which have found that amorphization or disruption of surface structure coincides with the modification of chemistry and can manifest as the formation of a `hard' graphitic surface layer  \cite{hofstetterQuantifyingDamageInduced2019,veghAPLNearsurfaceModificationPolystyrene2007, veghJAPMolecularDynamicsSimulations2008, gokanJESDryEtchResistance1983,zekonyteNIaMiPRSBBIwMaAInvestigationDrasticChange2005}.   
In IBS, sputter rates decrease rapidly after the removal of oxygen and nitrogen, which occurs in the beginning of the sputter process; the formation of the hard graphitic surface layer dramatically reduces the rate of ion removal from the material.

Because of the presence of Si in pV4D4, the effect of IBS treatment on the pV4D4 film differs remarkably from that observed for pHEMA and p4VP, both in terms of its wrinkling behavior (as shown in \figref{fig:AFM_Matrix}) and the modifications of its chemical structure. Although the surface of pV4D4 is likewise graphitized, an increased O content was detected on the pV4D4 surface.
The XPS spectra in \figref{fig:Total_C1sXPS}(c,f,i) show that, although the C-related peaks attributed to polymerized V4D4 disappear and graphitic C emerges, there remain pronounced higher binding energy peaks corresponding to CH$_3$ (C1), C–OH (C2), C–O (C3), and C=O (C4) species following IBS, as shown in \figref{fig:Total_C1sXPS}(f). This is shown by the reduced binding energy and a long asymmetric tail toward lower binding energies.
Additionally, the intensity of the graphitic peak and the accompanying increase in C1, C3, and C4 peaks after water vapor exposure subsequent to IBS in \figref{fig:Total_C1sXPS}(i) are relatively small, suggesting minimal change in stoichiometry due to the already large presence of O and C/O bonds. 

In HEMA, the ratio of C:O is 2:1, while in the V4D4 monomer, the ratio is 3:1.
One question that follows from this analysis is: `why does pV4D4 preserve O content after IBS treatment while pHEMA does not?'
There are a few factors that support an explanation for this question, depending crucially on the presence of Si. 
For pHEMA, it is likely that during IBS treatment, the oxygen leaves not as O$_2$ or atomic O, but as CO$_2$, lowering the O content even further than 2:1, whereas in pV4D4, O atoms released from their original bonds can continue to bond to amorphous Si. 
That atomic O in polymers could be removed from the surface during IBS as CO$_2$ was suggested by Kanski \textit{et al}. \cite{kanskiJPCLEffectOxygenChemistry2016}.
Furthermore, there is a significant difference in the bonding energies between C/O and Si/O depending on whether the bond is single or double; singly bonded O has a bond strength of 360 kJ/mol to C and 452 kJ/mol to Si, while double bonded O has strengths of 715 kJ/mol to C and 590 kJ/mol to Si \cite{normanPeriodicityPBlockElements1997,weinholdONatureSiliconOxygen2011}. This suggests that CO$_2$, which comprises C/O double bonds, is less energetically favorable to form compared to SiO$_2$, which comprises Si/O single bonds.
Additionally,  CO$_2$ has a Gibbs free energy of formation of $\Delta G_f^{\degree} = -394.4 \ \text{kJ/mol}$, while for SiO$_2$, the Gibbs free energy of formation is approximately $\Delta G_f^{\degree} = -856.4 \ \text{kJ/mol}$.
Hence, after the IBS treatment, the presence of Si stabilizes the O content in the film, even after it has been amorphized. 

\begin{table}[h!]
    \centering
    \resizebox{\columnwidth}{!}{%
    \begin{tabular}{c c c c c}
    \hline
         & Elements & No-IBS (\%) & IBS (\%) & IBS + H$_2$O (\%)\\
         \hline
    pHEMA & C  & 70 & 93 & 85\\
          & O  & 30 & 7  & 15\\
    \hline
          & C  & 86 & 91 & 27\\
    p4VP  & O  & 4  & 6  & 36\\
          & N  & 10 & 3  & 37\\
    \hline
          & C  & 62 & 63 & 62\\
    pV4D4 & O  & 32 & 28 & 29\\
          & Si & 6  & 9  & 9\\
    \hline
    \end{tabular}%
    }
    \caption{ Chemical composition of the three films as a function of processing step. pHEMA contains only C and O.  p4VP contains C, N, and impurity O. pV4D4 contains C, O, and Si in large amounts. Percentages are accurate to 2\%.}
    \label{tab:Polymer_XPS_Stoich}
\end{table} 

\subsection{FTIR Study of Bulk Film Chemistry for pHEMA}

\begin{figure}[H]
    \centering
    \includegraphics[width=\linewidth]{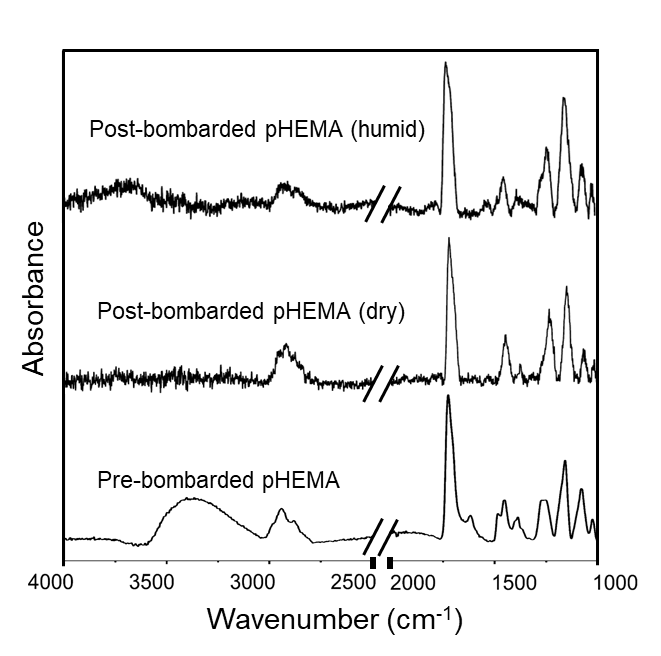}
    \caption{FTIR of IBS sputtered pHEMA film before and after exposure to a humid environment.}
    \label{fig:FTIR_HEMA}
\end{figure}

In \figref{fig:FTIR_HEMA}, we present the IR absorption spectra of the pHEMA films before IBS, after IBS, and after the sputtered film is exposed to a humid atmosphere.
In the spectrum in \figref{fig:FTIR_HEMA} taken before ion bombardment, a broad band centered at $\approx$ 3450 cm$^{-1}$ is observed; the O-H stretching from the hydroxyl groups is responsible for this band. These hydroxyl groups are from absorbed water \textit{prior} to bombardment. 
After bombardment, this band is strongly reduced in both dry and humid samples, indicating a reduction in hydroxyl groups after IBS treatment.
The C-H stretching region (3000 to 2850 cm$^{-1}$), which is related to CH$_2$ and CH$_3$ groups in the polymer backbone, is also strongly reduced after bombardment, suggesting a loss of hydrogen and damage to the aliphatic structure at the surface, as observed in XPS.
The ester carbonyl peak at $\approx$ 1720 cm$^{-1}$ is still present after bombardment in both dry and humid samples.
In the fingerprint region, clear peaks are seen before bombardment at $\approx$ 1450 cm$^{-1}$ (CH$_2$/CH$_3$ bending), $\approx$ 1270 cm$^{-1}$ (C-O stretching in ester), and $\approx$ 1080 cm$^{-1}$ (C-O stretching in alcohol and ester groups). 
After bombardment, these peaks become weaker in both spectra. 
This indicates the breakdown of ester side chains and alcohol groups. 
The weak band around $\approx$ 1640 cm$^{-1}$, which is related to water molecules or residual C=C from unreacted HEMA monomer, also changes after treatment. 
Between the two post-bombardment samples, the dry sample shows sharper and more visible peaks, while in the humid sample some bands appear broader and less clear.
This observation can be explained by the presence of absorbed water and hydrogen bonding in the modified polymer structure. 
Therefore, humidity exposure affects the FTIR spectrum in addition to the chemical changes caused by bombardment.
These absorption band assignments support the XPS peak assignments in \figref{fig:Total_C1sXPS}(g).

\subsection{\textit{In Situ} XPS Studies of Film Decomposition Under Sputtering}

\begin{figure}[h!]
    \centering
    \includegraphics[width=1\linewidth]{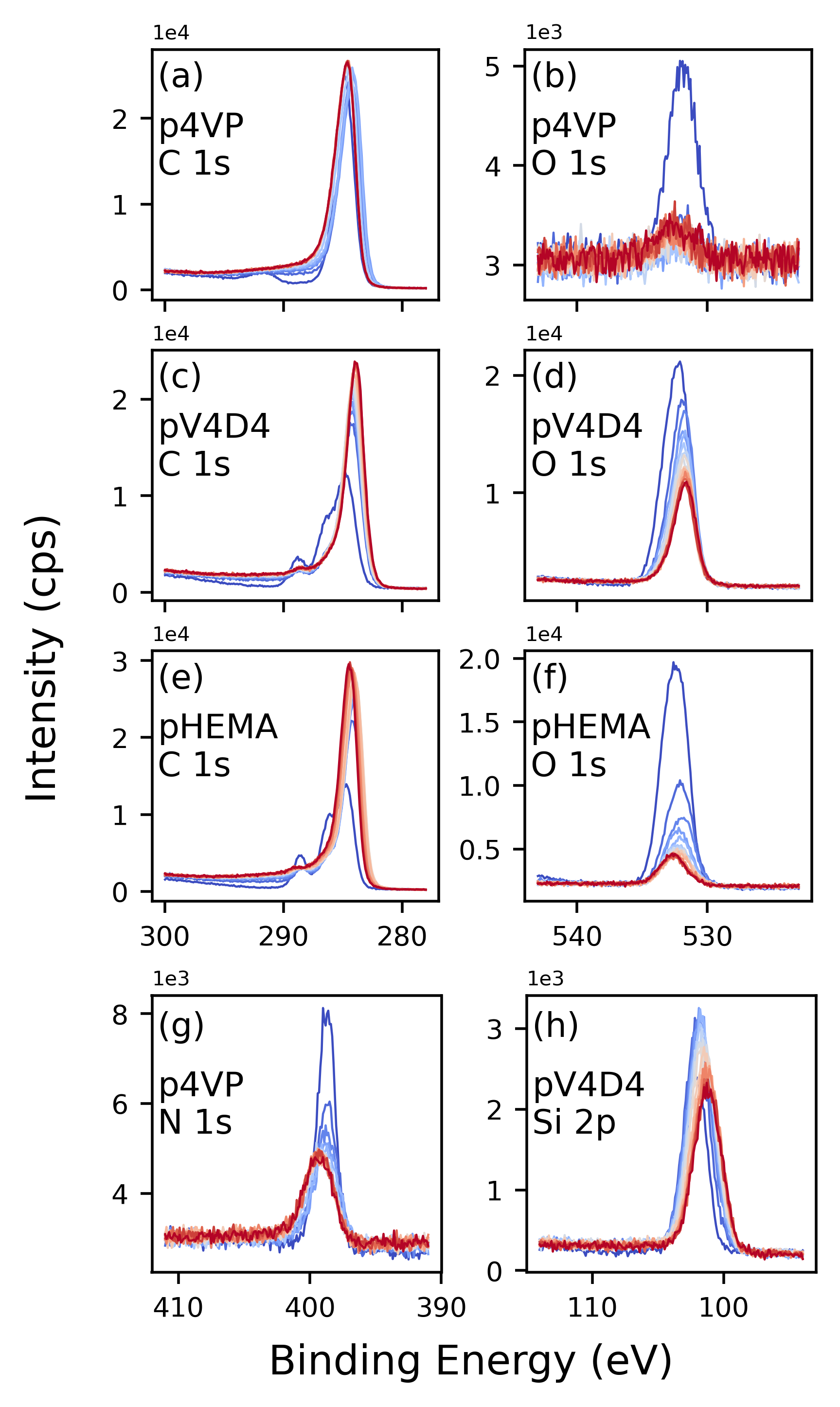}
    \caption{\textit{In situ} XPS of polymer films during IBS as a function of sputter step. C 1s spectra show graphitization, O 1s spectra demonstrate oxygen removal, and changes in N 1s and Si 2p spectra indicate amorphization on these sites. Color scheme corresponding to first scan pre-sputter (blue) and final scan (red).}
    \label{fig:XPS_Cascade}
\end{figure}

Given that wrinkle formation was observed at a humidity level ($\sim$ 60\% RH) that is comparable to normal ambient conditions, it suggests that the surface chemistry of the polymer films could change under normal room conditions after IBS. Consequently, \textit{ex situ} XPS may not accurately represent the true surface chemistry of polymers immediately after IBS. To address this potential limitation, \textit{in situ} XPS was performed to study the evolution of polymer surface chemistry during the IBS process.

\begin{figure}[h!]
    \centering
    \includegraphics[width=.8\linewidth]{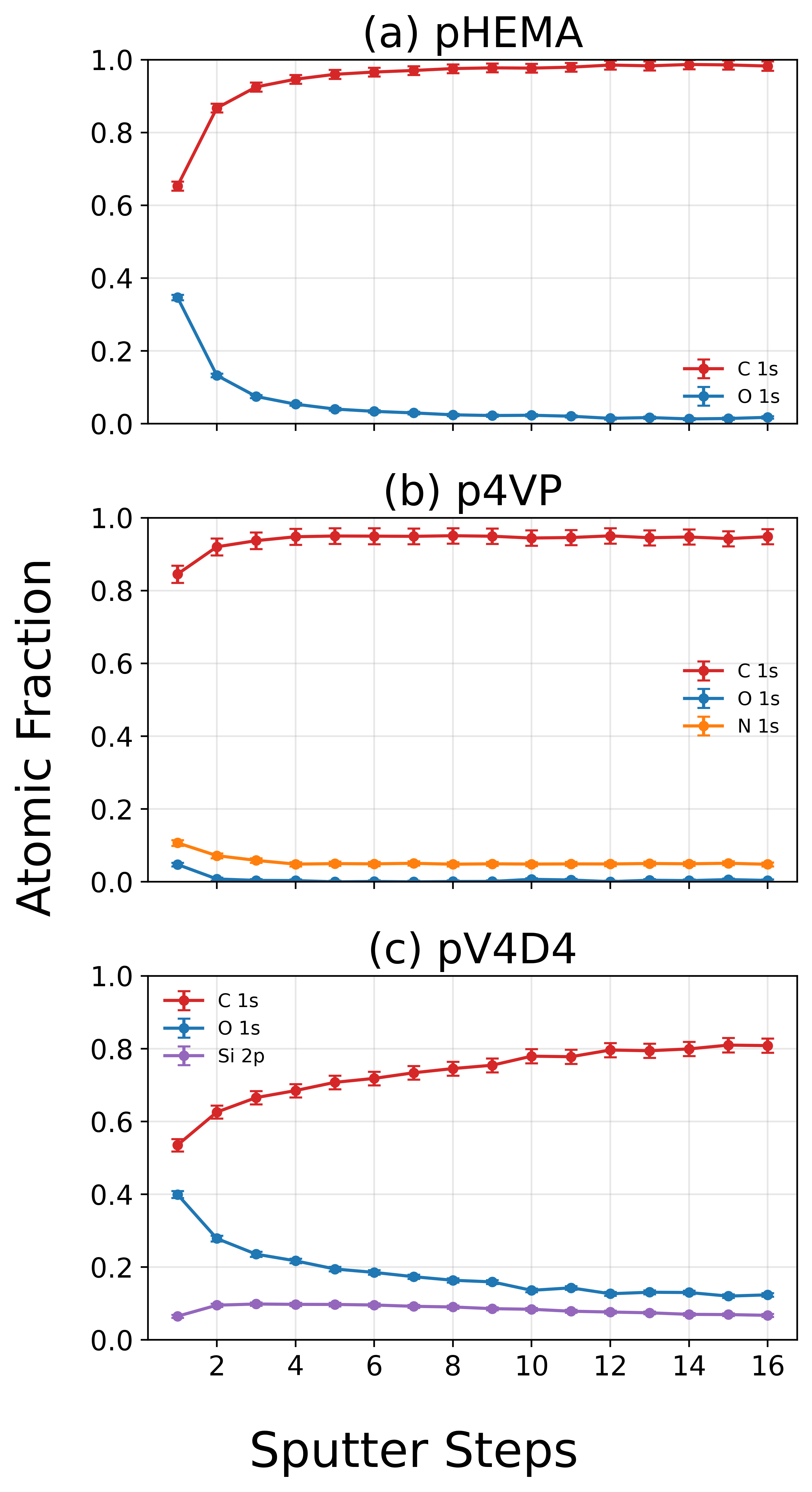}
    \caption{Stoichiometry of (a) pHEMA, (b) p4VP, and (c) pV4D4 as a function of sputter step at 2 keV $\text{Ar}^+$ of IBS from \figref{fig:XPS_Cascade}  }
    \label{fig:XPS_stoich}
\end{figure}

Using \textit{in situ} XPS with Ar IBS, we show the stepwise change of the polymer surface species prior to surface oxidation in \figref{fig:XPS_Cascade} with the corresponding changes in stoichiometry in \figref{fig:XPS_stoich}.
The changes in \figref{fig:XPS_Cascade} occur over the course of 16 sputter steps, with a total sputtering time of 96 s at an Ar energy of 2 keV. These changes are consistent with the \textit{ex situ} XPS data shown in \figref{fig:Total_C1sXPS}, demonstrating that there are no intermediate phases between the original polymer film and the graphitic surface phases formed as a result of IBS. The \textit{in situ} XPS shows that the complete change of the surface occurs within $\sim$ 2 min of sputtering ($\sim$\(2.9\times 10^{15}\) ions cm\(^{-2}\) in ion fluence) at 2 keV. In fact, from the stoichiometric evolution presented in \figref{fig:XPS_stoich}, after a sharp initial drop in oxygen, the films reach a relatively stable state within the first few sputter steps.

Modifications in chemical structure and composition (Fig.~\ref{fig:XPS_Cascade} and~\ref{fig:XPS_stoich}) occur rapidly and subsequently reach an apparent steady state, closely resembling the \textit{ex situ} XPS results in \figref{fig:Total_C1sXPS}. This indicates that the IBS treatment produces a chemically stable surface predominantly composed of graphitic or graphitized carbon species.

\section{Discussion}

\begin{figure}[ht!]
    \centering
    \includegraphics[width=\linewidth]{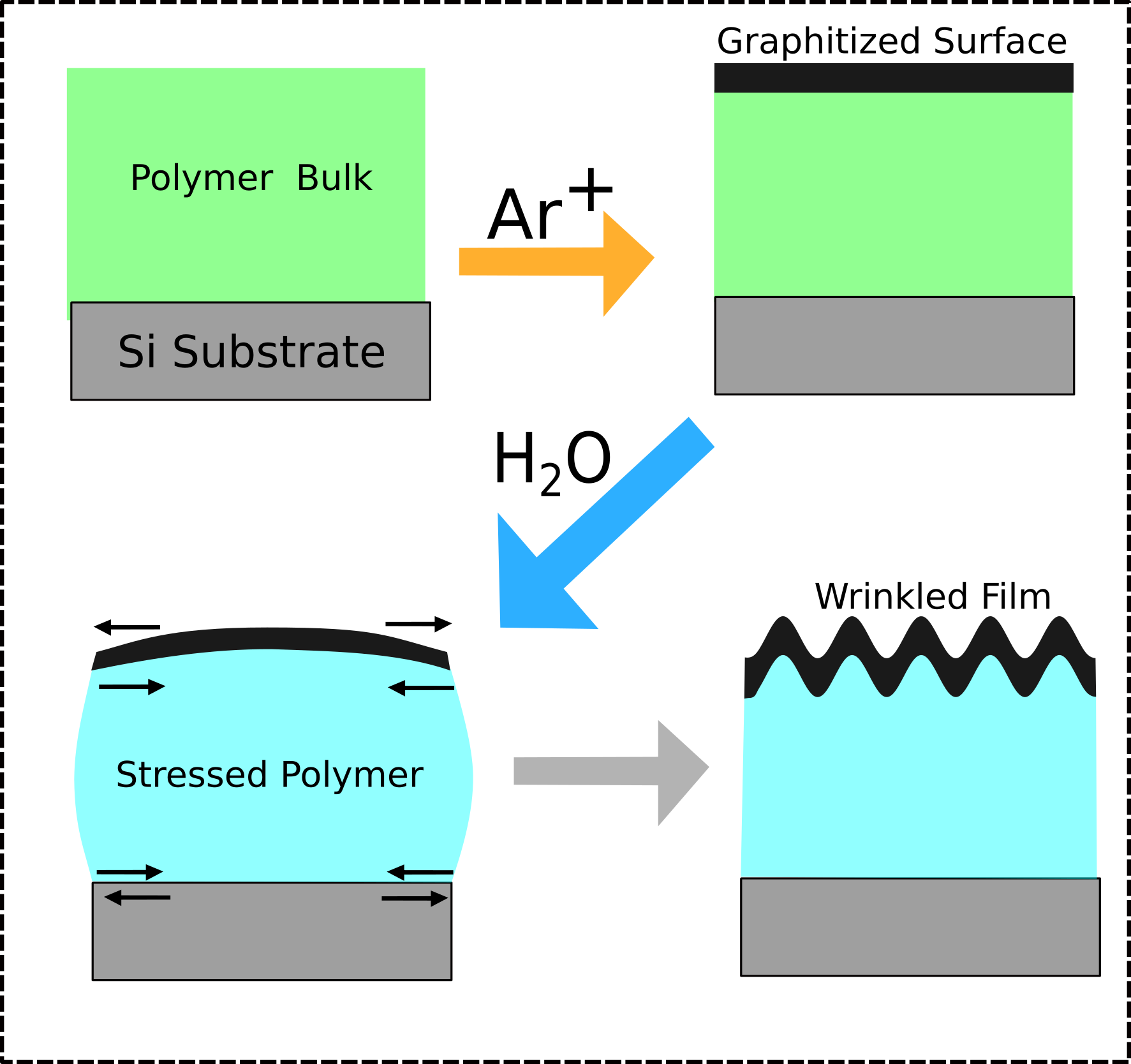} 
    \caption{Wrinkle formation as a result of IBS treatment followed by exposure to water vapor.
    After IBS treatment, a graphitic layer forms on the surface of pHEMA, p4VP, and pV4D4.
    As the pHEMA and p4VP films underneath the graphitic layer absorb water, they expand more than the graphitic surface layer, leading to the buildup of stress which results in wrinkling.}
    \label{fig:IBS_Wrinkle_Scheme}
\end{figure}

The IBS-induced wrinkling of PDMS films has been extensively investigated in previous studies \cite{moonPNASUWrinkledHardSkins2007,Jeong2015,jeongLTailoringOrientationPeriodicity2016,Arias2021}. In our present work, the study of wrinkling behavior is extended to a broader range of polymer thin films, specifically pHEMA, p4VP, and pV4D4 deposited on Si substrates. Of these three polymers, none developed wrinkles after IBS when kept in a dry atmosphere. Wrinkle formation was observed on pHEMA and p4VP films only after exposure to water vapor post-IBS at a relative humidity of approximately 60\% RH, while pV4D4 films remained topographically flat under the same humid conditions.

Both \textit{ex situ} and \textit{in situ} XPS demonstrate the graphitization of the surfaces of all three polymer films induced by the IBS treatment. Within the XPS sampling depth (10 nm), the evolution of the surface chemical structure and composition is observed to reach a steady state by the end of the IBS process. Considering that the penetration depth of 2 keV $\text{Ar}^+$ ions in polymers is less than 10 nm \cite{zekonyteNIaMiPRSBBIwMaAInvestigationDrasticChange2005,Delcorte2001}, it is therefore reasonable to infer that the rigid graphitized surface layer generated by IBS has a thickness below 10 nm, as schematically depicted in \figref{fig:IBS_Wrinkle_Scheme}.

The wrinkling behavior observed exclusively after both IBS treatment \textit{and} water vapor exposure resembles recent experimental findings by Ahmad \textit{et al}. \cite{ahmadAFMSurfaceWrinklingPlasmaExposed2025}, who reported that surface wrinkling in plasma-exposed PDMS films occurs only after water vapor sorption. In their work, plasma treatment produced an oxidized surface layer on the PDMS. Because the water sorption rate is higher in oxidized PDMS than in the underlying unmodified PDMS, the resulting swelling and in-plane expansion of the oxidized surface layer exceed those of the underlying PDMS film. This mismatch in expansion induces an in-plane stress in the oxidized surface layer which, in turn, gives rise to surface wrinkling. Since the oxidized surface layer induced by plasma exposure exhibits a greater volumetric expansion than the underlying unmodified film, the driving stress for surface wrinkling can be primarily ascribed to water sorption and the consequent swelling of the plasma-modified surface layer. It is natural to assume that the thickness of the underlying unmodified film remains constant over the duration of exposure to water vapor. Since the wrinkling wavelength is known to depend on the thickness of this underlying film \cite{Evensen2009,Hyun2009}, no time-dependent evolution of the wavelength is observed in the recent experimental study by Ahmad \textit{et al}. \cite{ahmadAFMSurfaceWrinklingPlasmaExposed2025}. However, our current results reveal a time-dependent evolution of the wavelength in both pHEMA and p4VP, indicating that the swelling and expansion of the modified surface layer are unlikely to cause the wrinkling behavior observed here.

As evidenced by the changes in the FTIR spectra demonstrating water uptake within the bulk pHEMA film, water vapor is capable of diffusing through the stiff graphitic surface layer generated by IBS and into the underlying unmodified polymer film. In the case of hydrophilic polymers such as pHEMA and p4VP, unmodified polymer films exhibit significant water absorption accompanied by swelling \cite{Gulsen2006,sahinerCAromaticOrganicContaminant2011}. In contrast, hydrophobic polymers such as pV4D4 exhibit low water absorption and, consequently, negligible swelling \cite{Reeja-Jayan2015}. Given the clear difference in the wrinkling behavior between the hydrophilic polymers, pHEMA and p4VP (which exhibit wrinkle formation), and the hydrophobic polymer, pV4D4 (which does not exhibit wrinkle formation), we infer that the swelling of the polymer film is the key driving force that gives rise to the surface wrinkling in this stiff graphitic surface layer/polymer film/Si substrate system. The volumetric expansion arising from water-absorption-induced swelling of the polymer film is constrained by the stiff graphitic surface layer and the hard Si substrate, thereby creating an in-plane compressive stress on the swollen polymer that leads to wrinkle formation (\figref{fig:IBS_Wrinkle_Scheme}). This underlying polymer film swelling mechanism has been previously examined in analogous systems comprising a stiff surface layer, a polymer interlayer, and a hard substrate \cite{jeongLTailoringOrientationPeriodicity2016,Chung2009,Vandeparre2010}. Among these studies, Vandeparre \textit{et al}. \cite{Vandeparre2010,Vandeparre2008} reported a similar evolution of the wrinkle wavelength and the wrinkle amplitude as a function of solvent vapor exposure time. The observed temporal dependence was attributed to solvent diffusion within the polymer and the viscous relaxation of the compressive stress generated by the stiff surface layer on the swollen polymers. 

From the perspective of water‐absorption–induced wrinkling in polymer films, our results demonstrate that water‐absorption–driven wrinkling following surface chemical modification by either plasma exposure or IBS is not restricted to PDMS, but can also occur in other polymers such as pHEMA and p4VP. We further hypothesize that, in the present systems (IBS‐modified pHEMA and p4VP films on a hard substrate), water absorption and the resulting swelling of the underlying polymer film beneath the modified surface layer constitute the primary driving mechanism for wrinkle formation. This contrasts with the mechanism proposed for PDMS by Ahmad \textit{et al}. \cite{ahmadAFMSurfaceWrinklingPlasmaExposed2025}, where the dominant contribution to wrinkling is attributed to the water-absorption-induced swelling of the modified surface layer itself. It is not unexpected that the present study yields results that differ from those reported by Ahmad \textit{et al}. \cite{ahmadAFMSurfaceWrinklingPlasmaExposed2025}, given the distinct chemical structures of PDMS, pHEMA, and p4VP, as well as the different nature of the modified surface layers (oxidized PDMS generated by plasma exposure versus graphitized pHEMA and p4VP produced by IBS). Together, these results demonstrate that the chemical structure of the starting polymer has an appreciable effect on the mechanisms and characteristics of wrinkle formation.

From the point of view of IBS, Jeong \textit{et al}. \cite{Jeong2015,Jeong2015Srep,jeongLTailoringOrientationPeriodicity2016} reported wrinkle formation on PDMS after IBS using a 600 to 2400 eV $\text{Ar}^+$ ion beam without requiring exposure to water vapor. They ascribe this IBS-induced wrinkling of PDMS to a similar interlayer swelling mechanism. The system is also described as comprising a stiff surface layer (chemically modified from PDMS by IBS), a swelling interlayer of unmodified PDMS, and a hard Si substrate. The swelling of the interlayer is attributed to the thermal expansion of the unmodified PDMS, wherein the heat originates from the conversion of the kinetic energy of $\text{Ar}^+$ ions into thermal energy \cite{Jeong2015,Jeong2015Srep,jeongLTailoringOrientationPeriodicity2016}. Although IBS-induced thermal expansion of polymer films in the present study cannot be completely ruled out, its contribution to wrinkle formation in the investigated polymers (pHEMA, p4VP, and pV4D4) appears to be negligible, or at least not the primary driving mechanism, as no wrinkling behavior was observed in any of these three polymers immediately after IBS treatment. Instead, this work introduces a different route to interlayer swelling following IBS, specifically swelling driven by water absorption. For hydrophilic polymers such as pHEMA and p4VP, ambient room humidity of approximately 60\% RH are sufficient to induce swelling after IBS. This finding implies that future investigations of IBS-induced wrinkling in polymer thin films must carefully account for the influence of ambient humidity when interpreting the resulting surface morphologies.

\section{Conclusion}
In this study, we demonstrated that IBS-induced surface wrinkling in polymer thin films can be strongly influenced by water absorption after surface chemical modification. pHEMA and p4VP films wrinkle only after IBS treatment followed by exposure to water vapor, whereas pV4D4 does not wrinkle under the same conditions. XPS and FTIR results indicate that IBS creates a graphitized surface layer and that subsequent water absorption in pHEMA and p4VP leads to swelling of the underlying polymer film. We therefore hypothesize that wrinkle formation in these systems arises from swelling that is mechanically constrained by the rigid silicon substrate and the stiff graphitized surface layer. In contrast, the absence of wrinkling in pV4D4 is attributed to its comparatively low water absorption and limited swelling response. These findings show that the chemical structure and hydrophilicity of the starting polymer have a significant effect on the mechanism and characteristics of wrinkle formation, and that ambient humidity must be carefully considered when interpreting IBS-induced surface morphologies in polymer thin films.

\section*{CRediT authorship contribution statement}
\textbf{Alessia Danagoulian:} Writing – original draft, Writing – review and editing, Visualization, Validation, Software, Investigation, Formal analysis, Data curation, Conceptualization. \textbf{Benli Jiang:} Writing – original draft, Writing – review and editing, Visualization, Validation, Software, Supervision, Project administration, Methodology, Investigation, Formal analysis, Data curation, Conceptualization. \textbf{Nicholas Russo:} Writing – original draft, Writing – review and editing, Visualization, Validation, Software, Supervision, Methodology, Investigation, Formal analysis, Data curation, Conceptualization. \textbf{Colette Abadie:} Writing – review and editing, Validation, Investigation, Formal analysis, Data curation. \textbf{Jalal Karimzadeh Khoei:} Writing – original draft, Writing – review and editing, Validation, Investigation, Formal analysis, Data curation. \textbf{Grace Pettis:} Writing – original draft, Writing – review and editing, Validation, Methodology, Investigation, Data curation, Conceptualization. \textbf{Jocelyn Zhang:} Writing – original draft, Writing – review and editing, Validation, Methodology, Investigation, Data curation, Conceptualization. \textbf{Neil Baker:} Writing – review and editing, Validation, Investigation, Data curation. \textbf{Eda Güney:} Writing – review and editing, Validation, Investigation, Data curation. \textbf{Omid Moradi:} Writing – review and editing, Validation, Investigation, Data curation. \textbf{Wei-Jing Chen:} Writing – review and editing, Validation, Investigation, Data curation. \textbf{Jiaqi Tang:} Writing – review and editing, Validation, Investigation, Data curation. \textbf{Robert Sims, Jr.:} Writing – review and editing, Validation, Investigation, Data curation. \textbf{Kevin E. Smith:} Writing – review and editing, Resources, Methodology, Conceptualization. \textbf{Gozde Ozaydin Ince:} Writing – original draft, Writing – review and editing, Validation, Supervision, Resources, Project administration, Methodology, Conceptualization. \textbf{Karl F. Ludwig, Jr.:} Writing – original draft, Writing – review and editing, Validation, Supervision, Resources, Project administration, Methodology, Funding acquisition, Conceptualization.

\section*{Declaration of competing interest}
The authors declare that they have no known competing financial interests or personal relationships that could have appeared to influence the work reported in this paper.

\section*{Acknowledgments}
This work was partly supported by the National Science Foundation (NSF) under Grant No. DMR-2117509 and PHY-2244795. We also thank the NSF Grant No. CHE-2216008 that purchased the PHI Genesis XPS. We gratefully acknowledge Ellie Stonecipher and Marlene Ludwig for their valuable technical assistance in establishing and optimizing the experimental setup for IBS. We acknowledge Boston University Chemistry Department Chemical Instrumentation Center and Serge Zdanovich for using FTIR.

\section*{Data availability}
The data that support the findings of this study will be made available upon reasonable request.

\newpage

\appendix
\section{AFM Characterization and Numerical Analysis of the Wrinkles}\label{subsec:AFM-analysis}
\begin{figure}[ht!]
    \centering
    \includegraphics[width=1\linewidth]{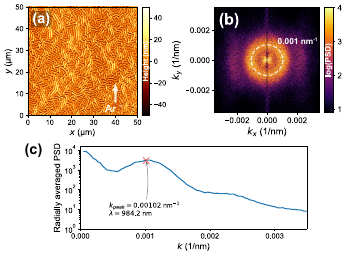}
    \caption{AFM image (a) and the corresponding 2D PSD (b) of wrinkled p4VP surfaces after 7 days in the humid atmosphere after IBS. The 1D PSD (c) is obtained by performing a radial average of the 2D PSD. The characteristic wrinkle wavelength is subsequently determined from this 1D PSD.}
    \label{fig:AFM_wavelength}
\end{figure}

To determine the characteristic wrinkle wavelength, the 2D PSD of the surface morphology was extracted from the AFM image, as \figref{fig:AFM_wavelength} shows. The spatial frequency (\(k\)) of the PSD corresponds to the inverse of the lateral scan length scale (\(L\)) in AFM, such that \(k = 1/L\). Since the wrinkle patterns are isotropic, the corresponding 2D PSD presents a ring-shaped feature. This ring-shaped feature correlates with the wrinkle, and the radius of the ring corresponds to the wrinkle wavelength. By performing a radial average of the 2D PSD, a one dimensional (1D) PSD is obtained (\figref{fig:AFM_wavelength} (c)). The peak position in the 1D PSD corresponds to the characteristic wrinkle wavelength. To determine this peak position, the spectral peak is fitted with a Gaussian function, and the wrinkle wavelength (\(\lambda\)) is obtained from the location of the maximum of the resulting Gaussian fit (\(k_{peak}\)). Thus, the wavelength is given by \(\lambda = 1/k_{peak}\).

The overall uncertainty of the characteristic wrinkle wavelength (\(\sigma_\lambda\)) mainly arises from two contributions: the Gaussian fitting procedure (\(\sigma_{fit}\)) and the spatial resolution limit of the PSD (\(\sigma_{res}\)). \(\sigma_{fit}\) is the standard error of the fitted result (\(k_{peak}\)). \(\sigma_{res}\) originates from the finite lateral scan length of the AFM measurement (\(L_{scan}\)). As a result, the resolution in reciprocal space for the PSD is limited to \(\Delta k = 1/L_{scan}\). Assuming that the probability of finding the peak position within this resolution limit (\(\Delta k\)) is uniformly distributed, the corresponding uncertainty is \(\sigma_{res}=\Delta k / \sqrt{12}\). Then, the combined uncertainty of the wrinkle wavelength in reciprocal space (\(\sigma_{k_{peak}}\)) is obtained by \(\sigma_{k_{peak}}=\sqrt{\sigma_{fit}^2+\sigma_{res}^2}\). Finally, the overall uncertainty of the characteristic wrinkle wavelength in real space (\(\sigma_\lambda\)) is derived by standard error propagation, yielding \(\sigma_\lambda= (\lambda\cdot\sigma_{k_{peak}})/k_{peak}\).

For the results presented in \figref{fig:AFM_Analysis}, each data point is derived from multiple AFM measurements at different lateral scan sizes. While the uncertainties of the wavelength were introduced previously, the uncertainty of the RMS roughness for each individual AFM measurement is assumed to be 5\% of the measured RMS value, which is attributed to the instrumental uncertainty of the AFM. All measurements contributing to a single data point in \figref{fig:AFM_Analysis} are subsequently combined by calculating a weighted mean, with the associated uncertainties determined through standard error propagation.

\begin{figure}[ht!]
    \centering
    \includegraphics[width=0.9\linewidth]{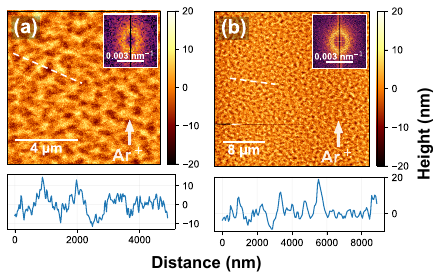}
    \caption{AFM images and the surface height profile along the the white dash line of wrinkled pHEMA surfaces after 6 days in the humid atmosphere after IBS. The scan sizes of the AFM images are $10\times10$ $\mu$m$^2$ for (a) and $30\times30$ $\mu$m$^2$ for (b), respectively. The inset presents the 2D PSD of the surface morphology.}
    \label{fig:AFM_HEMA}
\end{figure}

\begin{figure}[ht!]
    \centering
    \includegraphics[width=0.9\linewidth]{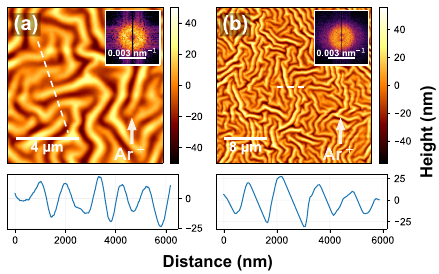}
    \caption{AFM images and the surface height profile along the the white dash line of wrinkled p4VP surfaces after 9 days in the humid atmosphere after IBS. The scan sizes of the AFM images are $10\times10$ $\mu$m$^2$ for (a) and $30\times30$ $\mu$m$^2$ for (b), respectively. The inset presents the 2D PSD of the surface morphology.}
    \label{fig:AFM_4VP}
\end{figure}

\pagebreak
\section{XPS Peak Models}\label{subsec:XPS-analysis}
In this study, we developed different lineshapes for each unsputtered pristine polymeric species. 
The lineshapes for pHEMA and p4VP were constrained by imposing fixed relationships among the corresponding peak areas. In contrast, for pV4D4 such constraints could not be applied, owing to the multiplicity of ways for V4D4 to polymerize. 
The carbon labels `C1, C2...' of the unsputtered species shown in Table \ref{tab:XPS_Lineshapes} correspond to the labels in \figref{fig:Molecules}.

The Gaussian-Lorentzian Product (GLP) used in this study is defined according to a recent article by Major \textit{et al}. \cite{majorS&IADetailedViewGaussian2022}. 
The model for the graphitic peak structure was a modified finite asymmetric Lorentzian function convolved with a Gaussian function, where the finite asymmetric Lorentzian was defined by:

\begin{equation}
    LF(x; x_0 , F,A, \alpha,\beta, w) = \begin{cases} 
          [L(x;x_0, F,F \frac{\pi}{2})]^{\alpha'} & x \leq x_0 \\
          [L(x;x_0, F,F \frac{\pi}{2})]^{\beta'} &  x  >  x_0 \\ 
       \end{cases}
\label{eqn:LF_a}
\end{equation}

The coefficients $\alpha'$ \& $\beta'$ are defined by:
\begin{equation*}
   \alpha= a_S - \frac{a_s-\alpha}{1 + 4 (\frac{x-x_0}{w})^2} 
\end{equation*}
with the asymmetry factor taken to be $a_s=3$. 
A finite asymmetric Lorentzian line shape was used because it tapers to a finite integrated area, which facilitates quantitative analysis while still representing the extended asymmetry arising from the metallic character of the graphitic phase \cite{gengenbachPracticalGuidesXray2021,majorJoVS&TAPracticalGuideCurve2020}.
For our graphitic carbon model, we obtain good agreement by choosing $\alpha=1;\beta=0.5$ and allowing $F,w$ and the FWHM of the convolving Gaussian function to vary. 
A pHEMA sample sputtered in vacuum was used to determine the lineshape, and the resultant fit has $F_L=0.24 \pm0.01$, $w = 200 \pm 22$, and $F_G = 0.40 \pm 0.01$, with a binding energy of 284.0 eV.
This model was then used as a composite fit component 
in fitting the \textit{in situ} XPS data presented in \figref{fig:XPS_Cascade}. 

\begin{table*}[h!]
    \centering
    \begin{tabular}{|c|c|c|c|c|c|c|}
\hline 
 Component & Binding Energy  (eV) & FWHM (eV)  & Parameters   & Lineshape Function      & Constraints  \\ 
 \hline
 pHEMA  &           &          &            &       &              \\\hline
    AdC &   284.3   &   1.95   &    $m = 0.2$ &  GLP     &             \\\hline  
    C1  &   285.26   &  1.24    &   $m = 0 $  &  GLP     &  $A= A_1$            \\\hline
    C2  &   285.94   &  1.08    &   $m = 0$   &  GLP     &  $A= 0.5 A_1 $           \\\hline
    C3  &   287.00  &    1.62    &   $m = 0$   &  GLP     &  $A= A_1$            \\\hline
    C4  &   289.05   &  1.48    &   $m = 0 $  &  GLP     &  $A= 0.5 A_1 $           \\
    
\hline 
  p4VP  &           &          &            &          &            \\\hline
    C1  & 284.24  &  1.61  &$m = 0.8 $    &  GLP &     $A=2 A_3$     \\\hline
    C2  & 285.04   &  1.51  &  $m = 0.65 $ &  GLP &    $A=2 A_3$       \\\hline
    C3  & 285.54   &  1.51  &  $m = 0.65 $ &   GLP & $A= A_3$           \\\hline
    C4  & 285.94   &  1.51  &  $m = 0.65 $ &  GLP &     $A=2 A_3$       \\ \hline
    $\pi^*$  &292.15 & 3.41 &  $m = 1 $    &   GLP &               \\
\hline 
 pV4D4  &           &          &            &             &        \\\hline
    C1  &   283.5  &   2.00   &     $m = 0.3$       &     GLP     &   \\\hline
    C2  &   284.5  &   1.68   &     $m = 0.3$       &     GLP     &   \\\hline
    C3  &   286.2  &   1.80   &     $m = 0.3$       &     GLP     &   \\\hline
    C4  &   288.5  &   1.48   &     $m = 0.3$       &     GLP     &   \\
    \hline
    \end{tabular}
    \caption{Table of XPS lineshapes used in the analysis. $m = 0$ indicates a complete Gaussian and $m = 1$ indicates a fully Lorentzian peak.  }
    \label{tab:XPS_Lineshapes}
\end{table*}
\FloatBarrier

 \bibliographystyle{elsarticle-num} 
 \bibliography{citation-combined}

\end{document}